# Kinetics and direct imaging of electrochemically formed palladium hydride for efficient hydrogen evolution reaction

Luca Camuti[∥,#,§], Se-Ho Kim[†,‡,§,*], Filip Podjaski[∥,¶,§,*], Miquel Vega-Paredes[†], Andrea M. Mingers[†], Tolga Acartürk[∥], Ulrich Starke[∥], Bettina V. Lotsch[∥,#], Christina Scheu[†], Baptiste Gault[†,⌐], Siyuan Zhang[†,*]

[∥]Max Planck Institute for Solid State Research, Heisenbergstraße 1, 70569 Stuttgart, Germany

[#]Department of Chemistry, University of Munich (LMU), Butenandtstr. 5-13, 81377 Munich, Germany

[†]Max Planck Institute for Sustainable Materials, Max-Planck-Straße 1, 40237 Düsseldorf, Germany

[‡]Department of Materials and Engineering, Korea University, Seoul 02841, Republic of Korea

[¶]Department of Chemistry, Imperial College, London, SW7 2AZ, United Kingdom

[⌐]Department of Materials, Royal School of Mines, Imperial College, London, SW7 2AZ, United Kingdom

[§]these authors contributed equally

[*]corr. Authors: sehonetkr@korea.ac.kr | f.podjaski@imperial.ac.uk | siyuan.zhang@mpie.de

**Abstract**

Active and reliable electrocatalysts are fundamental to renewable energy technologies. $PdCoO_2$ has recently been recognized as a promising catalyst template for the hydrogen evolution reaction (HER) in acidic media thanks to the formation of active $PdH_x$. In this article, we monitor the transformation of single $PdCoO_2$ particles during HER, and confirm their almost complete transformation to $PdH_x$ with sub-millimeter depths and cracks throughout the particles. Using *operando* mass spectrometry, Co dissolution is observed under reductive potentials, leading to $PdH_x$ formation, whereas the dissolution partial current is found to be 0.1 % of the HER current. The formation of $PdH_x$ is confirmed through secondary ion mass spectrometry and quantitatively analyzed by atom probe tomography, enabled by isotope labelling of hydrogen using heavy water. Despite dry storage and high vacuum during sample preparations, an overall composition of $PdD_{0.28}$ is measured for the $PdH_x$ sample, with separation between alpha- (D-poor) and beta- (D-rich) $PdH_x$ phases. The $PdH_x$ phase formed on $PdCoO_2$ particles is stable for a wide electrochemical potential window, until Pd dissolution is observed at open circuit potentials. Our findings highlight the critical role of a templated growth method in obtaining stabilized $PdH_x$, enabling efficient HER without the commonly slow activation processes observed in Pd. This offers insights into the design of more efficient electrocatalysts for renewable energy technologies.

**Keywords:** Palladium hydride, electrocatalyst activation, hydrogen evolution reaction, hydrogen detection, atom probe tomography

**Introduction**

Electrolytic water splitting into hydrogen and oxygen is one of the most promising technologies to transform and store intermittent renewable energy as chemical fuels.[1] However, given the thermodynamics of water splitting (1.23 V at room temperature), minimizing the catalytic overpotential and increasing the stability often require the use of precious-metal catalysts.[2] This is especially the case for modern proton exchange membrane (PEM) electrolyzers, which can operate with flexible load, but need harsh acidic conditions.[3] However, the high cost of precious-metal catalysts like Pt and Ir requires that they present an outstanding and stable hydrogen evolution reaction (HER) activity for their successful large-scale commercial implementation.[4]

$PdCoO_2$ was recently shown as a precursor to highly efficient HER[5, 6] and water oxidation electrocatalysts in acidic media. Single crystals of $PdCoO_2$ shows an extraordinarily long electron-mean-free path and remarkably high in-plane electrical conductivity (30% larger than elemental Cu), minimizing resistive losses.[7, 8] The HER performance of $PdCoO_2$ improves with time in acidic media, which is attributed to the surface transformation to a Pd-based layer by leaching out $[CoO_2]^{2-}$. Based on its electrochemical characteristics, the HER-active surface layer was hypothesized to be $PdH_x$, which was found to be isotropically expanded by 2 % compared to bulk Pd due to a strain-templated growth method.[5, 6]

Besides the superior electrochemical performance, there is little known about the electrochemically formed $PdH_x$ on the $PdCoO_2$ template. For reliable electrosynthesis, it is important to understand the kinetics of the surface transformation, as shown by "How fast?" in Scheme 1. Moreover, the stability of the $PdH_x$ ("How stable?") is of prime importance for its HER application. Critically related is understanding of "How much?" hydrogen is incorporated into the layer, and whether they are homogeneously distributed. Furthermore, the extent of the transformation ("How deep?" into PdCoO$_2$ template) is essential to allow optimization of material use and longevity of the electrocatalyst.

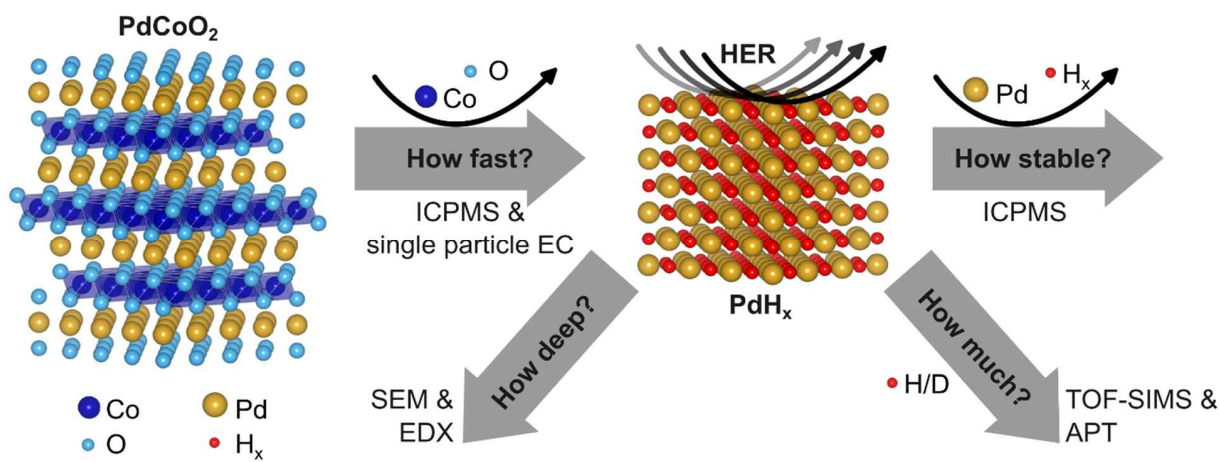

**Scheme 1.** Open questions on the electrochemically formed PdH$_x$ layer from PdCoO$_2$ template and the experimental techniques used to address them: Inductively-coupled plasma mass spectrometry (ICPMS), scanning electron microscopy (SEM), energy dispersive X-ray spectroscopy (EDX), time of flight (TOF) – secondary ion mass spectrometry (TOF-SIMS), and atom probe tomography (APT).

In this article, state-of-the-art characterization tools are applied to address these open questions. We show the extent of transformation towards PdH$_x$ by long-term single particle experiments and the kinetics of Co and O dissolution by *operando* techniques. The electrochemically formed PdH$_x$ remains stable under HER conditions, and exhibits high activity even after exposure to vacuum, bypassing need for an additional activation period. The stable Pd hydride phase is confirmed by two independent mass spectrometry techniques after using a deuterated electrolyte, as isotope label enabling indirect but precise quantification of the hydrogen uptake into our Pd with expanded lattices.

**Results**

The surface transformation of $PdCoO_2$ to an highly active catalyst during HER operation has been demonstrated.[5, 6] There, tens of nanometers of $Pd(H_x?)$ were observed on the surface, which can grow to a few hundred nanometers with extended HER operation.[5] Compared to the volume of $PdCoO_2$ powder particles (μm to mm scale), the transformation appeared superficial. Here, we investigate the transformation of individual particles to correlate it with the observed HER current, removing the averaging effects of multi-particle electrodes. To quantify electrolyte-assisted phase transformation to a hydride phase, we used Ar saturated electrolytes to ensure the comparability to the deuterium labeled spectroscopic measurements, and to avoid the incorporation of ambient $H_2$. Hence, we reference potentials against Ag/AgCl ($V_{Ag/AgCl}$, saturated KCl), since the $H_2$ pressure during the experiments is unknown and subject to change during the activation process.

The crystallites change their morphology during the long-term chronoamperometry under HER conditions (−0.35 $V_{Ag/AgCl}$, 1 M Ar-saturated $H_2SO_4$ electrolyte), as shown by SEM micrographs in Figure 1. Before HER activation (Figure 1a), the $PdCoO_2$ particle had smooth crystal surfaces dominated by basal-plane (0003) facets. The surface area can be estimated well by the SEM image in this case to normalize the measured current. The projected area in Figure 1a was determined as 0.817 $mm^2$. During the first 70 h of HER operation, the catalytic activity kept increasing until the current stabilized at ~ −0.85 $A/cm^2$ (Figure 1b). Note that despite having only one particle, the collected current (−7 mA) is 4 orders of magnitude higher than a bare C-paste electrode (Figure S1). Smaller crystallites had even faster activation processes occurring on all electrolyte-exposed facets, reaching plateaus of current densities close to −1 $A/cm^2$ after less than 10 h (Figure S2-S5). Note that in all of our single particle measurements, the current density approaches −1 $A/cm^2$, which is the targeted current density range for commercial PEM electrolysis.[9, 10]

After the initial activation, a high stability in current density is observed in all our measurements. It is also noteworthy that even after the measurements were interrupted, i.e., the electrode was removed from the electrolyte environment, dried and stored in vacuum, the high current density was immediately reached by re-applying the HER conditions (Figure 1b). Furthermore, no similar activation of a pure Pd wire toward the high-activity PdHx phase is observed under comparable conditions (Figure S6).

The surface transformation of single particles after HER operation was examined using SEM. As shown in Figures 1c, d, S7, and S8, the catalyst kept its size and shape, but the surface became rough with many cracks forming, effectively increasing the surface area and contributing to increasing currents. A cross-section of the particle aged for 140 h of HER operation, Figure 1d and S9, shows that the PdH$_x$ covers ~90 % of the imaged area and extends over 100 µm below the surface. This leaves only small pockets of PdCoO$_2$ around the center (Figure 1d). Taking the transformed volume into account, 0.08 % of the observed current go into Co (see SI Section 1.2). EDX analysis (Figure S10–S13) further reveals very little Co (Table S1, Co/Pd ~ 0.06) in the transformed Pd-based region.

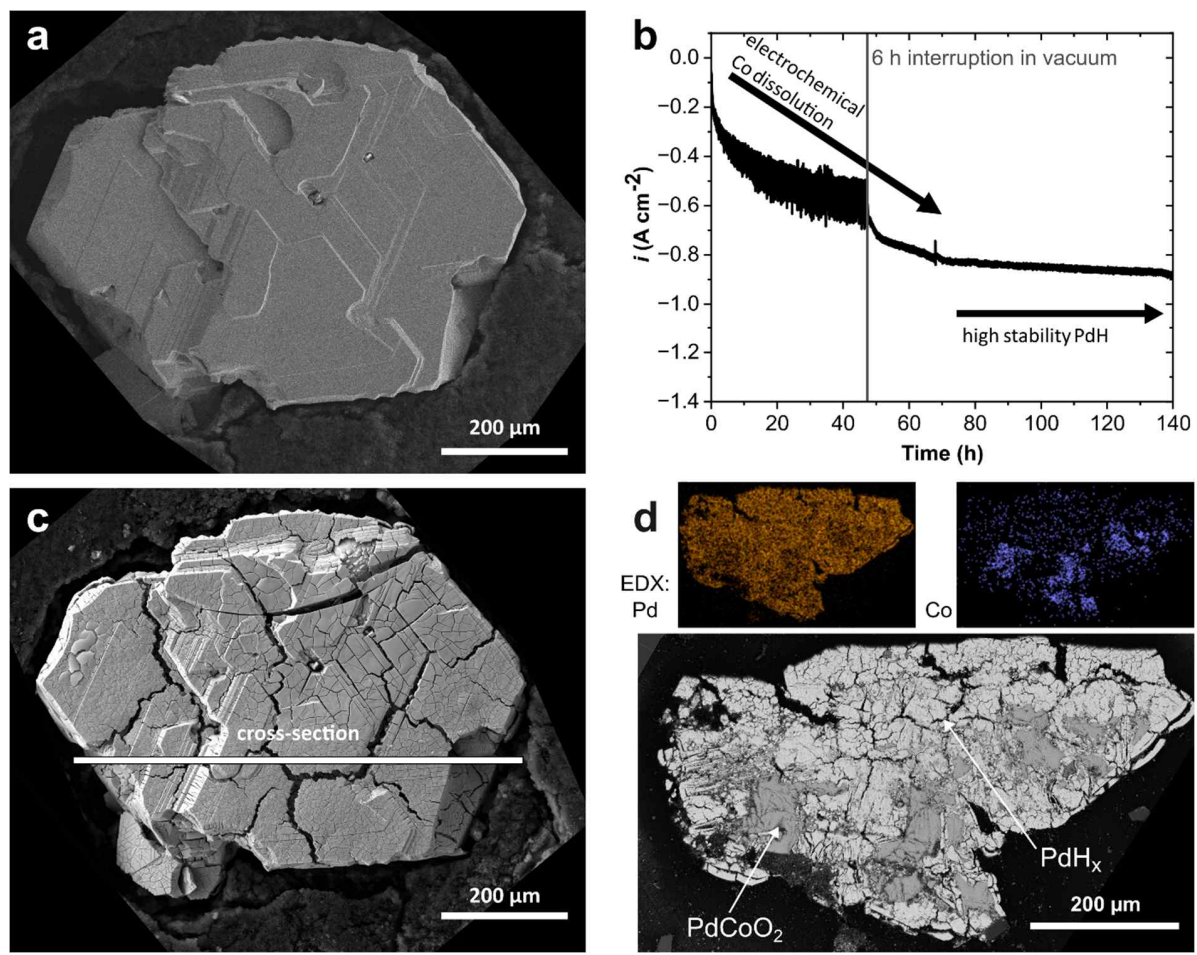

**Figure 1.** Analysis of the electrochemical aging behavior of PdCoO$_2$. The tests were performed in Ar bubbled 1 M H$_2$SO$_4$ for 140 h at a potential of −0.35 V$_{Ag/AgCl}$. (a) SEM picture of the pristine PdCoO$_2$ crystal on a carbon paste electrode. (b) Geometric current density observed during the aging process of 140 h. The geometric surface area of the catalytically active crystals was obtained from SEM pictures. Strong bubble formation leads to fluctuations of the effective current. The electrochemistry was interrupted for 6 h after 48 h to obtain SEM images (Figure S11). (c) SEM image of the same crystal as in (a) after catalysis. The approximate cut line for the cross-section in (d) is indicated. (d) EDX maps and backscattered electron image of a cross-section of the aged crystal. Darker shades of grey correspond to PdCoO$_2$, lighter shades of grey resemble the PdH$_x$.

Having demonstrated stable HER after almost complete transformation from $PdCoO_2$ to $PdH_x$, ICPMS measurements were conducted to investigate the kinetics of Co dissolution from pristine $PdCoO_2$ particles. O is also leaching out during the experiments, but cannot be quantitatively analyzed by our set-up. By coupling the ICPMS with a scanning flow cell (SFC),[11] the setup allows for *operando* examination of the corrosion products in the electrolyte (SI Section 1.3). The high sensitivity of the SFC-ICPMS enables quantification of even trace amounts of dissolution from electrocatalysts[12-16], photocatalysts[17-19], and in particular, noble metals.[20, 21]

We examined the transformation of pristine $PdCoO_2$ particles via two electrochemical protocols, cyclic voltammetry (CV) scans into the HER potentials (Figure 2a), and galvanostatic holds under HER conditions (Figure 2b). Both methods lead to the formation of $PdH_x$ from $PdCoO_2$.[5, 6, 22] To facilitate the discussion and comparison of electrochemical potentials with literature data, the reversible hydrogen electrode ($V_{RHE}$, derived from the Ag/AgCl reference electrode assuming sufficient local hydrogen saturation *operando*, details in SI) scale is given in brackets. Multiple $PdCoO_2$ particles were dispersed on a fluorine-doped tin oxide electrode, while the current density was normalized by the geometric area of the SFC, 1 $mm^2$.

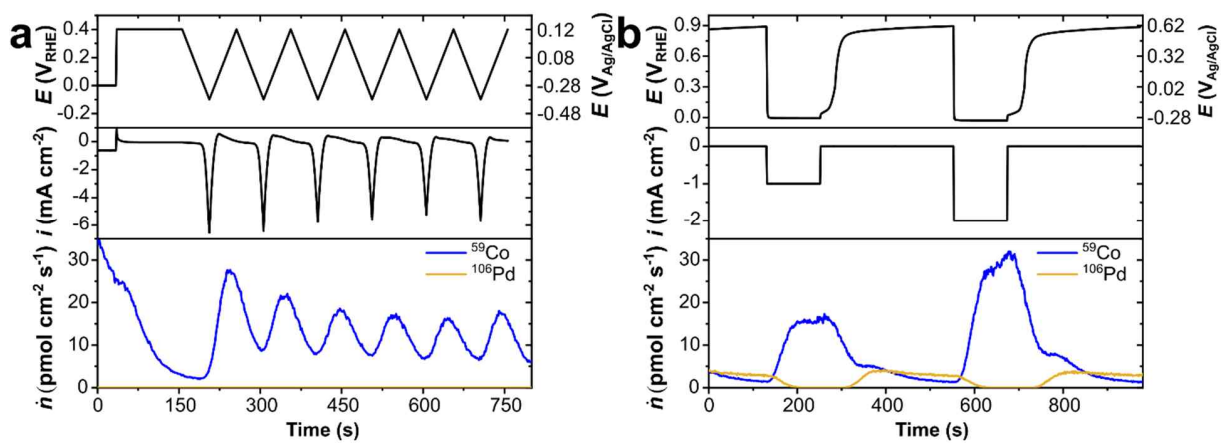

**Figure 2.** Dissolution properties and kinetics of PdCoO$_2$ in 0.1 M H$_2$SO$_4$ under *operando* conditions: (a) cyclic voltammetry scans at 10 mV/s, (b) galvanostatic holds at OCP, −1, and −2 mA/cm$^2$. Time profiles of the electrode potential, geometric current density, as well as Co and Pd dissolution rates are displayed in the top, middle and bottom rows, respectively.

As shown in Figure 2a, Co dissolution from pristine PdCoO$_2$ is observed as soon as the working electrode comes into contact with the electrolyte within the SFC. With the potential kept anodic to HER, the dissolution rate progressively decreases. Such transient dissolution behavior is commonly observed and can be attributed to off-stoichiometric compositions on the surface.[17] As the CV cycles between 0.12 and −0.38 V$_{Ag/AgCl}$ (0.4 and −0.1 V$_{RHE}$) are repeated, Co dissolution shows a pattern that matches the CV periodicity, while the amount of Pd dissolution remains negligible (Figure 2a). Within each CV cycle, the Co dissolution starts increasing at each cathodic scan to HER potentials. From the third CV cycle onward, the amount of Co dissolution stabilizes. The Pourbaix diagrams show that at acidic pH and around 0 V$_{RHE}$, solid PdH$_x$ and aqueous Co$^{2+}$ are the thermodynamically stable species.[23] Therefore, we propose the following redox reaction to account for the dissolution of PdCoO$_2$, parallel to the HER.

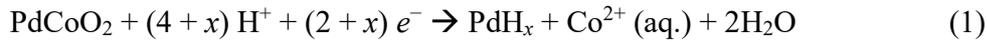

PdCoO$_2$ + (4 + $x$) H$^+$ + (2 + $x$) $e^-$ → PdH$_x$ + Co$^{2+}$ (aq.) + 2H$_2$O     (1)

For each unit of PdCoO$_2$, one electron (those not being active for HER) reduces Co$^{3+}$ to Co$^{2+}$, and 1 + $x$ electrons reduces Pd$^+$ and protons to form PdH$_x$. H absorption into Pd is well known under HER conditions;[24] the concentration $x$ will be discussed later. As the CV scan reverses back to anodic potentials, Co dissolution reaches a maximum at ~0 V$_{Ag/AgCl}$, which is likely due to delayed release of reduced Co species on PdCoO$_2$ to the electrolyte flow.

The kinetics of the transformation process is further explored by monitoring Co and Pd dissolutions from pristine PdCoO$_2$ under galvanostatic conditions at 0 (open circuit potential (OCP) measurement), −1, and −2 mA/cm². As shown in Figure 2b, Co dissolution is proportional to the current density of the electrode. The partial current of the dissolution reaction $i_{Co}$ can be evaluated based on the detection rate of $Co^{2+}$, $\dot{n}_{Co}$:

$$i_{Co} = \dot{n}_{Co} \times (-1) \times 96485 \text{ C/mol} \tag{2}$$

At $i_{total} = -1$ mA/cm², the partial current is $i_{Co} \approx -1.5$ µA/cm². At $i_{total} = -2$ mA/cm², the partial current is $i_{Co} \approx -3$ µA/cm². The branching ratio of $i_{Co} / i_{total} \approx 0.15\%$ underlines that HER is the predominant contribution to the total current (HER Faraday efficiency close to unity),[5] while the proportional scaling indicates that the Co leaching current is kinetically coupled to the HER.

Although Co and O continuously dissolve during HER, Pd remains stable against dissolution, as shown in Figure 2a. This is consistent with the conservation of overall shape and volume of the crystal as $PdCoO_2$ is transformed to $PdH_x$ (Figure 1). Our SFC measurements further reveal the stability range of the electrochemically formed hydride phase. As shown in Figure 2b, Pd dissolution is also negligible during HER at −1 or −2 mA/cm². However, once the HER stops, the electrode OCP drifts back to anodic potentials (0.5~0.6 $V_{Ag/AgCl}$) within minutes. Pd dissolution was recorded from ~0.5 $V_{Ag/AgCl}$, and plateaued as OCP stabilizes at ~0.6 $V_{Ag/AgCl}$ (~0.9 $V_{RHE}$). At these OCP conditions, the Pd dissolution rate eventually exceeds that of Co leaching. This finding is consistent with reports that Pd dissolves in acids at potentials anodic of 0.9 $V_{RHE}$.[21] Our results confirm that such high anodic potentials, which are reached at OCP conditions of electrochemically obtained Pd-based material, are to be prevented in practical cases to avoid catalyst losses. It is worth noting that the instability at OCP conditions has been found for many HER catalysts, such as Pt, $MoS_2$.[12, 16, 25, 26] To protect the electrochemically formed $PdH_x$ from dissolution, the electrode needs to be maintained under HER working conditions or well below the dissolution threshold of 0.5 $V_{Ag/AgCl}$ at pH = 1.

After determining the electrochemical stability of the active Pd hydride material, its composition is analyzed to understand the content and distribution of hydrogen. Firstly, we employed time-of-flight-secondary ion mass spectrometry (TOF-SIMS) measurements to analyze a sample after 15 min activation at HER conditions (−0.3 $V_{Ag/AgCl}$). The short-term activation was implemented to enable the detection of $PdH_x$ formed on the sample surface, while keeping the surface $PdH_x$ layer thin enough to be completely removed during the measurement process, exposing the pristine $PdCoO_2$ crystallite. Since hydrogen is ubiquitous in the environments and hence its origin is difficult to clarify, we used deuterium as a hydrogen indicator. Therefore, 1 M $D_2SO_4$ prepared from concentrated $H_2SO_4$ and $D_2O$ was used as electrolyte during electrochemical aging (see SI Section 1.2).

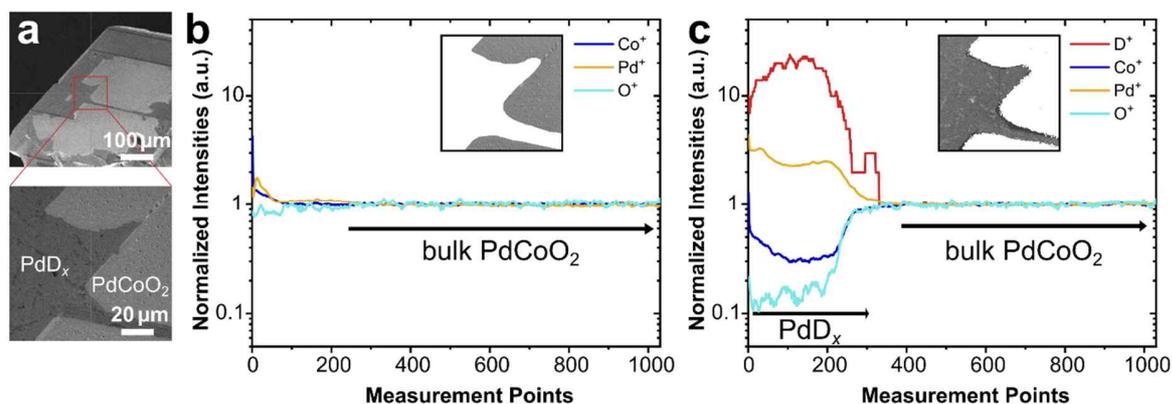

**Figure 3.** TOF-SIMS measurements performed on electrochemically aged $PdCoO_2$ samples (15 min at −0.3 $V_{Ag/AgCl}$). (a) Secondary electron image of the aged $PdCoO_2$ to be analyzed, with brighter and darker areas highlighting different sample regions (pristine and transformed material, respectively). Close-up shows analysis area. (b) TOF-SIMS data obtained from the area shown in the inset, associated with $PdCoO_2$ resulting from an area of delaminated capping layer. The data for SIMS-generated $Pd^+$, $Co^+$ and $O^+$ is averaged over 10 points. The $D^+$ channel detected singular events associated with random noise and was hence excluded from the plot. The signal intensity was normalized to the intensity associated with bulk $PdCoO_2$, indicated by the arrow. (c) TOF-SIMS data obtained from the area shown in the insert, associated with $PdD_x$. The same data processing as in (c) was performed. The measurement points are proportional to the measuring depth. For pristine PdCoO$_2$ the proportionality factor is 0.9±0.2 nm per measurement point.

As shown in Figure 3a, a 500 × 500 μm$^2$ planar region of the crystal was selected by secondary electron imaging for the TOF-SIMS measurement. Figure 3a shows the measurement region (100 μm$^2$) containing both the PdD$_x$ phase (darker shade of grey) and an area with exposed PdCoO$_2$ surface (light gray, exposed by mechanical peel-off of the capping layer). TOF-SIMS profile obtained from bulk PdCoO$_2$ (Figure 3b) shows constant Pd:Co:O stoichiometry and negligible amounts of D$^+$ throughout its depth (Figure S14). In contrast, the TOF-SIMS profile from the PdD$_x$ layer (Figure 3c) reveals a significant amount of D$^+$ on the surface, decreasing with depth (down to 300 measurement points). This correlates with the increase in Pd counts, as well as the reduction of Co and O signals, confirming the presence of a PdD$_x$ phase. As shown in Figure S15, the depth profile of a 1 h aged sample exhibits signals from PdD$_x$ and PdCoO$_2$ comparable to the sample after 15 min HER activation, although the depth of the PdD$_x$ region is, as expected, 3–4 times thicker. Moreover, the surface after 1 h aging is completely covered by PdD$_x$, exhibiting cracks similar to long-term aged crystals. Due to the lack of calibration standards and complex excitation processes of secondary ions leading to varying detection efficiencies, a precise quantification of D$_x$ using TOF-SIMS is not possible.

In order to quantify the amount of hydrogen/deuterium in PdH$_x$/PdD$_x$ and gain insight into their spatial distribution, we used atom probe tomography (APT) to image deuterium in 3D at sub-nanometer resolution.[27-29] As shown in Figure S16, pristine PdCoO$_2$ without electrochemical treatment shows a uniform distribution of Pd, Co and O close to their theoretical stoichiometry. A description of the specimen preparation including images of the needle prepared by a gallium focused ion beam (FIB) can be found in the supporting information (SI Section 1.6 and Figure S17).

As residual hydrogen stemming from the APT chamber can cause a high background of hydrogen and complicate the quantification,[28, 30] we used deuterated samples for analysis (similar to the TOF-SIMS analysis). After 1 h of HER activation of $PdCoO_2$ in 1 M $D_2SO_4$, APT specimens were prepared from different areas of the surface: activated $PdD_x$ region (Pd layer), exposed $PdCoO_2$ surface ($PdCoO_2$ bulk), and regions in between (interface) (Figure 4a). As shown in Figure 4b, progressively fewer Co and O atoms are detected, as we approach the activated $PdD_x$ layer, which is consistent with the observed Co dissolution.

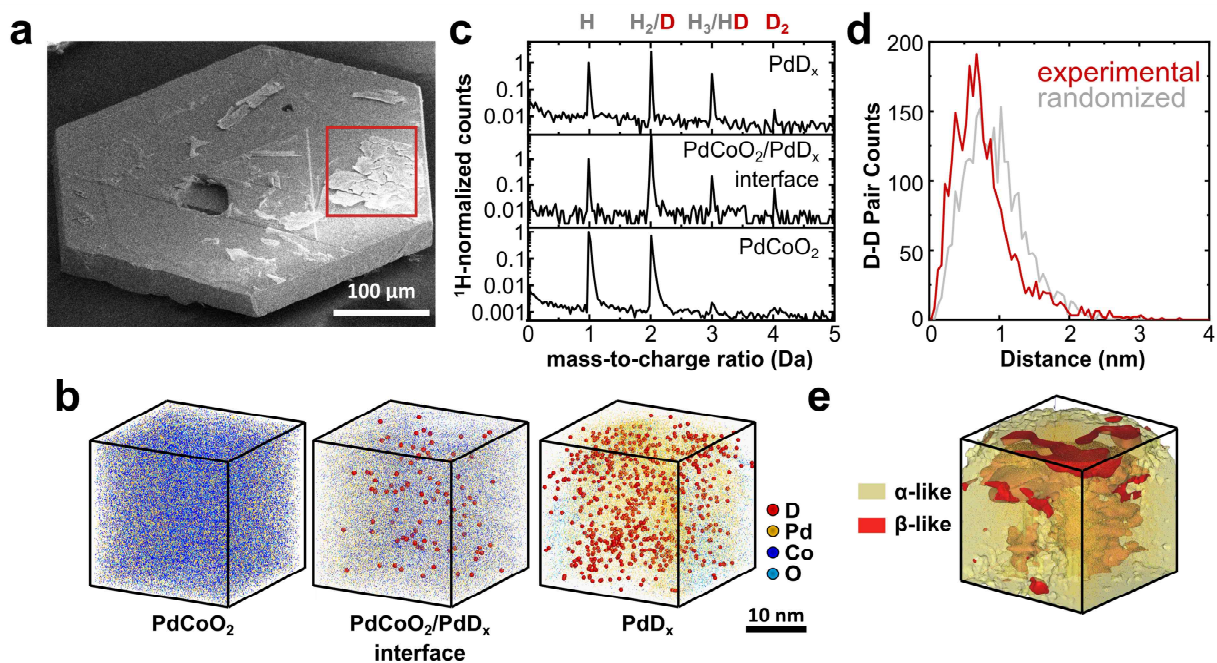

**Figure 4.** Characterization of $PdD_x$ nanocrystals on $PdCoO_2$ after 1 h HER activation in deuterated electrolyte (1.0 M $D_2SO_4$). (a) SEM image showing areas of $PdD_x$ layer on top of $PdCoO_2$, areas scrapped off of Pd layer ($PdCoO_2$ bulk), and interfaces between them. (b) 3D atom maps and (c) integrated mass spectra for H, D, DH, $D_2$ detection of the APT samples taken from the 3 regions: $PdCoO_2$ bulk, $PdCoO_2/PdD_x$ interface, and $PdD_x$ layer. (d) D-D pair analysis and

(e) separation of volumes with α-like (PdD$_{0.08}$) and β-like (PdD$_{0.64}$) compositions in the PdD$_x$ layer sample. 20 % deuterium was used as cut off.

In the mass spectra from three datasets (PdCoO$_2$ bulk, PdCoO$_2$/PdD$_x$ interface, and PdD$_x$ layer), significant amounts of D, DH, and D$_2$ were observed (Figure 4c). Spurious H$_2$ species generated during the APT measurement due to the electric field may influence the analysis of D.[30, 31] Therefore, to confirm the presence of D, the charge-state ratio of Co and H/D was compared for different pre/post HER steps (Figure S19). The increase in the relative D signals by a factor of 20 after HER evidences that the signal of D is not from the gaseous H$_2$ but rather the pre-occluded D inside the Pd.

By integrating all the D-related peaks (D$^+$, DH$^+$, D$_2^+$) in Figure 4c, the average PdD$_x$ phase composition from the Pd layer dataset in Figure 4b was determined to be $x \approx 0.28$. The homogeneity of D distribution was analyzed by evaluating the frequency of D-D atoms to form nearest-neighbor pairs. As shown in Figure 4d, D shows a tendency to segregate as compared to a randomized distribution at $x \approx 0.28$. The iso-compositional surface of D at 20 at.% (corresponding to PdD$_{0.28}$) separates the sample into D-rich and D-poor volumes. As shown in Figure 4e, the D-rich volume is mostly at the inner core of the sample, and it has an overall composition of PdD$_{0.64}$. According to the Pd-H phase diagram,[32] this composition is close to the lower limit of β-PdH$_x$ at room temperature (PdD$_{0.5\sim0.6}$), so that the D-rich region is hence named as β-like volume. On the other hand, the α-like volume mostly locates at the outer shell and has an overall composition of PdD$_{0.08}$. It is noteworthy that Co signals were found in both α-like and β-like volumes, both at Co/Pd ratios of ~0.1. The APT data confirm that the electrochemically formed PdD$_x$ has coexistence of α-phase and β-phase compositions. It is also likely that a much bigger or even the whole volume had a β-PdD$_x$ composition during electrochemical charging, and hydrogen desorption during dry storage and sample preparation at high vacuum (< $10^{-5}$ mbar) [33] resulted in an outer shell that is depleted in D.

## Discussion

1. Fast formation of $PdH_x$ phase

In this study we investigated the formation behavior of $PdH_x$ obtained by electrochemical Co and O dissolution form $PdCoO_2$. We looked into the composition of the *operando* formed phase to confirm the hydride phase.

Here, we found that only 0.15 % of the initially observed current is attributed to the dissolution of Co, happening in parallel to the HER. The activity stabilizes when the dissolution of surface near Co with an average rate of 0.08 % is complete, forming a highly active $PdH_x$ layer. This activity is maintained over up to 140 h of HER and is not much affected by the removal of the electrolyte or the exposure to vacuum, since the hydrogen stays bound stably in the hydride phase. We confirmed the formation of the hydride phase by deuterium labeling and quantified the $PdD_x$ phase at $x \approx 0.28$ *ex-situ*, after vacuum treatment.

The observed dissolution of Co and O at cathodic potentials is well known for cobalt oxide species in acidic electrolytes,[34, 35] and documented in the Pourbaix diagram.[23] The applied negative potentials stabilize the *operando* formed hydride layer against dissolution akin to electro-deposited Pd-nano particles.[36] The transformation towards $PdH_x$ starts immediately and continues until either complete or a thick, dissolution rate-limiting $PdH_x$ layer is formed. The differences in branching ratio for the long-term HER and the SFC experiments are therefore due to a higher, initial dissolution rate, tapering off and approaching zero when the transformation is complete.

PdCoO$_2$-derived PdH$_x$ is activated orders of magnitudes faster than nanocrystalline Pd by CV.[5, 37] PdH$_x$ obtained from PdCoO$_2$ reaches high, Pt like, HER activity after 1000 CVs (0 to –0.1 V$_{RHE}$)[5] compared to nanocrystalline Pd which needs 130 000 CV cycles (0.3 to –0.45 V$_{RHE}$) to activate to a similar degree.[37] Our experiments showed no activation of metallic Pd under similar conditions for up to 60 h (Figure S6). We explain this by the significant activation barrier for hydrogenation in Pd, since the lattice mismatch between α-PdH$_x$ ($x < 0.015$) and β-PdH$_x$ ($x > 0.6$) is significant (from 3.89 to 4.04 Å).[38, 39] As such, non-template based Pd requires significant time to form active β-PdH$_x$.[38, 40] Hence, we conclude that *immediate* activation is associated to, and enabled by the templating effect of the Pd-sublattice in the delafossite structure in PdCoO$_2$ (Pd-Pd distance: 2.83 Å), which is very close to β-PdH$_x$ (2.84 Å), leaving the grown PdH$_x$ material in this lattice-expanded state even if detached from the substrate.[5, 38, 39]

2. Enhanced HER activity of PdH$_x$ phase on PdCoO$_2$ compared to Pd

This stable hydride enables a constant modification of the HER mechanism on PdCoO$_2$, as indicated by the change in Tafel slope (from $76 \pm 13$ to $38 \pm 3$ mV/dec),[5] similar to pre-hydrogenated PdH$_x$ with $x > 0.6$ and a Tafel slope < 40 mV/dec.[41, 42] It is in stark contrast to nanocrystalline and bulk Pd, where the activation is very sluggish or does not happen at all under similar conditions (Figure S6), leading to smaller changes in Tafel slope (between 96 and 119 mV/dec), indicative of the discharge reaction as the rate-limiting step of bulk Pd (Volmer Step, Tafel slope of ~100 mV/dec).[36, 37, 43] The observed high activity of the PdH$_x$ phase stems from the modified surface catalytic mechanism, as in a lattice expanded hydride, hydrogen adsorption is not rate limiting any more as on Pd. Although previous results suggested a lower intrinsic, surface normalized activity (j$_0$) for β-PdH$_x$ due to its tensile strain and the consequently changed d-band position, those findings appears to be canceled by operando surface roughening, and would not affect the electro-catalytic mechanism, as discussed previously.[5, 36, 44-46] Our dissolution process therefore offers a fast and facile alternative pathway to established methods of PdH$_x$ nanocrystallite formation,[47, 48] without the requirement of prolonged hydrogenation, and with crucial stability benefits.[37, 49]

3. Stability of PdH$_x$ phase and its ability to retain H

Since the *operando*-formed active PdH$_x$ phase stays stable even after a full transformation took place, the stability of the β-PdH$_x$ phase appears independent of the template. Stopping the HER does not lead to an activity degradation of the *operando*-formed active material, even if the electrolyte is removed from the system and if the sample is exposed to vacuum). However, electrolyte contact and relaxation towards the OCP can result in Pd dissolution, as seen in the SFC measurements (Figure 2b).

The D/Pd ratio of 0.28 cannot be explained by bulk phase diagrams of hydrogen (or deuterium) and palladium, since this composition resides within the miscibility gap between α-PdH$_x$ and β-PdH$_x$.[50] In bulk α-PdH$_x$ the H-content can reach 0.009 (α-PdD$_x$, $x \leq 0.01$),[50, 51] however in nanocrystalline Pd-H systems an expansion of the α-phase is observed. This is due to its nanocrystalline nature, hydrogen trapping can be taking place at grain boundaries, allowing for higher H content in the α-phase, reaching H/Pd ratios of up to 0.12 for particles larger than 1.4 nm.[52, 53] With the observed ratio, an expansion of the α-phase seems unlikely, even when additional, kinetically stabilized hydrogen[38, 39, 54] or grain boundaries and defects in the *operando* grown layer are taken into account, which allow for more trapping sites than in individual, uniformly sized Pd nanocrystals.[55]

The observed composition meets the minimum solubility ($x \geq 0.28$) of the β-PdH$_x$ phase in nanocrystals.[52] However, the presence of two distinct phases within the Pd layer – α-and β-phase – was clearly observed in the APT data, suggesting that the electrochemically grown Pd layer supports phase co-existence with D segregation contributing to the formation of domains. Such phase separation is consistent with the thermodynamic miscibility gap and may reflect the influence of nanoscale effects on hydrogen solubility and distribution. Especially, since the FIB sample preparation and APT conditions can lead to hydrogen desorption, the observation of a β-phase region supports the *operando*-formation of a highly active β-PdH$_x$ phase.

4. Outlook: Design of template-based HER catalysts

As such, it appears that strain engineering and hydrogen retention, as observed herein, are both crucial factors for enabling fast and efficiently operating electrocatalysts, especially in cases where hydrogen absorption leads to activation.[56] Both are likely related, since complex phase transformations typically requiring high amounts of reorganization energy, are to be minimized to ensure fast material performance/response and little energy losses. Facilitating the hydrogen sorption behavior when it is slow and catalytically rate limiting in pristine materials appears to be a design principle for novel catalyst materials with persisting activity and stability. This includes designing precursor materials as templates which form the active phase *operando*. Such materials may be of value for various processes with rate-limiting intermediate steps relying on adsorption, including other cases of catalysis (water oxidation, $CO_2$ or nitrogen reduction).[57-61] Facilitated hydrogen uptake, enabled by expanded lattices, may also be beneficial for other purposes:

Hydrogen storage and transport in metal hydrides, where thermodynamically and kinetically facilitated hydrogen absorption, transport and release are key, will benefit from this approach.

Going beyond the deliberate incorporation of hydrogen species into active catalyst materials, the question of the role of hydrogen in other catalytic systems arises. Understanding the role of *in-situ* absorbed hydrogen species in catalytic process could help to understand the activity of other classical catalysts as well as support the development of novel catalyst concepts - especially since the detection of hydrogen in hydrogen rich environments is difficult and hence the formation of such could be missed in many instances.

**Conclusion**

To summarize, the transformation process of $PdCoO_2$ template into the HER electrocatalyst $PdH_x$ is characterized by *operando* and *ex-situ* measurements. Electrochemistry on individual $PdCoO_2$ particles shows stable HER performance at current densities close to −1 A/cm², even after the entire particle is almost completely transformed to $PdH_x$. The transformation process is characterized by the dissolution of Co, which is kinetically related to HER at a branching ratio of ~0.1%, leading to the formation of a $PdH_x$ layer on top of the $PdCoO_2$ template. The incorporation of hydrogen into Pd is analyzed by SIMS and APT using deuterium labeling. $PdD_x$ formation is only detected in areas that show Co leaching. The averaged D content is determined to be $x = 0.28$. The deuterium content of the *operando*-formed $PdD_x$ falls within the miscibility gap of bulk Pd, suggesting a separation into α- and β-phase on the nano-scale. We attribute some loss of deuterium with the APT sample preparation and associate the active phase with nano-particular β-$PdH_x$.

Our investigations also clarify practical application aspects. Open circuit conditions positive of 0.8 $V_{RHE}$ should be avoided during interrupted HER operation, as Pd dissolution was identified in this region. The transformed catalyst shows comparable activity to bulk platinum, and operates without the need for reactivation when exposed to ambient conditions, which hinders other Pd-based catalysts from being useful in electrolyzer contexts. Therefore, $PdH_x$ derived from $PdCoO_2$ is a viable alternative to Pt based catalysts, especially for variable load scenarios, due to its high stability and immediate activity without the need for reactivation. Developing Pd as an HER catalyst would effectively double the availability of possible catalyst materials, since Pt and Pd have similar abundances.[62] Thin-film or nano-chemistry approaches could lead to even faster activation while increasing the cost efficiency of $PdCoO_2$ as catalyst-precursor material by minimizing the amount of catalyst per unit surface area. We anticipate that the findings herein inspire further research on strained structures, which can enable faster or more stable hydrogen uptake and storage for tailored electrocatalytic fuel production, or for engineering sorption and transport properties of hydrogen and its isotopes.

**Acknowledgements**

We acknowledge the support from the FIB and APT facilities at MPIE by Uwe Tezins, Andreas Sturm, and Christian Broß. B.V.L. and L.C. acknowledge supported by the Center for Integrated Quantum Science and Technology (IQST), the Cluster of Excellence e-conversion (Grant No. EXC2089) and the Max Planck Society. S.-H.K. acknowledges supports from National Research Foundation of Korea (NRF) funded by Ministry of Science (RS-2024-00450561, ICT) and KIAT-MOTIE (RS-2024-00431836, Technology Innovation Program). F.P. acknowledges UKRI funding under the grant reference EP/X027449/1. B.G. acknowledges financial support from the German Research Foundation (DFG) through DIP Project No. 450800666. S.Z. acknowledges funding from the DFG under the framework of the priority program 2370 (project number: 502202153).